\documentclass[twocolumn]{aastex631}

\usepackage{graphicx}	
\usepackage{amsmath}	
\usepackage{hyperref}
\usepackage{cprotect}

\newcommand{\jr}{\color{black}}

\allowdisplaybreaks

\begin{document}

\title{Fleeting but not Forgotten: the Imprint of Escaping Hydrogen Atmospheres \\ on Super-Earth Interiors}

\author[0000-0001-7615-6798]{James G. Rogers}
\affiliation{Department of Earth, Planetary, and Space Sciences, The University of California, Los Angeles, 595 Charles E. Young Drive East, Los Angeles, CA 90095, USA}

\author[0000-0002-0298-8089]{Hilke E. Schlichting}
\affiliation{Department of Earth, Planetary, and Space Sciences, The University of California, Los Angeles, 595 Charles E. Young Drive East, Los Angeles, CA 90095, USA}

\author{Edward D. Young}
\affiliation{Department of Earth, Planetary, and Space Sciences, The University of California, Los Angeles, 595 Charles E. Young Drive East, Los Angeles, CA 90095, USA}



\begin{abstract}
Small, close-in exoplanets are divided into two sub-populations: super-Earths and sub-Neptunes. Most super-Earths are thought to have lost their primordially accreted hydrogen-dominated atmospheres via thermally driven winds. We consider the global chemical equilibrium of super-Earths and the lasting impacts of their fleeting hydrogen atmospheres. We find that hydrogen is efficiently sequestered into the interior, oxidising iron and endogenously producing $\sim0.5-1.0\%$ water by mass. As the atmospheres of super-Earths are continuously sculpted by mass loss and chemical equilibration, they remain hydrogen-dominated by mole (number) fraction but become steam-dominated by mass, which may be observable with \textit{JWST} for planets transitioning across the radius valley. One of the main effects of efficient sequestration of hydrogen into the interior is to produce an under-dense bulk interior compared to that of Earth. We predict bulk densities of super-Earths to be $\sim 5.0 \text{~g~cm}^{-3}$ for a $1M_\oplus$ planet, which is consistent with high-precision mass measurements and also population-level inference analyses from atmospheric escape models. 
\end{abstract}



\section{Introduction} \label{sec:intro}

Hydrogen plays a central role in controlling various stages of planetary formation and evolution. For the population of small, close-in exoplanets \citep[super-Earths and sub-Neptunes e.g.][]{Howard2012,Fressin2013,Silburt2015,Mulders2018,Zink2019}, hydrogen-dominated gas is initially accreted from a planet's nascent protoplanetary disc to form a primordial atmosphere \cite[e.g.][]{Lee2014,Ginzburg2016}. As the disc disperses, much of this atmosphere can be lost through a boil-off stage, in which the disc rapidly drains onto its host star, inducing extreme atmospheric escape \citep{Owen2016,Ginzburg2016,Rogers2024}. Then, once the planet receives direct irradiation from its host star, its remaining hydrogen-dominated atmosphere is bombarded by stellar irradiation, inducing further atmospheric escape via X-ray/EUV (XUV) photoevaporation \citep[e.g.][]{Owen2013,LopezFortney2013} and core-powered mass-loss \citep[e.g.][]{Ginzburg2018,Gupta2019,Rogers2024}.

The loss, or retention, of hydrogen-dominated atmospheres has been used to explain various features in the exoplanet demographics accurately. Examples include the position and slope of the observed radius gap \citep[e.g.][]{Fulton2017,VanEylen2018,Petigura2022} as a function of orbital period and stellar mass \citep[e.g.][]{Gupta2019,Gupta2020,Rogers2021,Rogers2021b}, the planet mass-radius diagram \citep[e.g.][]{Lopez2014,ChenRogers2016,Kubyshkina2022,Rogers2023b}, as well as the short orbital period Neptune-desert \citep{OwenLai2018}. It is also known that many sub-Neptunes require a significant hydrogen atmosphere to reproduce their observed bulk densities \citep[e.g.][]{JontofHutter2014,Weiss2014,Benneke2019}. Direct observational evidence also exists for the prevalence of hydrogen-dominated atmospheres in the form of Ly-$\alpha$ and H$\alpha$ transit spectroscopy \citep[e.g.][]{DosSantos2023a}, as well as recent atmospheric characterisation of sub-Neptunes with \textit{JWST} \citep[e.g.][]{Madhusudhan2023,Wogan2024}.

A less well-explored avenue of investigation, however, is the potential impact of hydrogen-dominated atmospheres on the bulk interiors of super-Earths and sub-Neptunes \citep[e.g.][]{Kite2016,Chachan2018,Olson2018,Kite2019,Kite2020,Lichtenberg2021,Kite2021,Schlichting2022,Misener2023,Suer2023,Charnoz2023}. Multiple works have shown that, for hydrogen-dominated atmospheres to explain the small planet demographics, the bimodal sub-populations of super-Earths and sub-Neptunes originally began as a single population of `Earth-like' interiors hosting hydrogen-dominated atmospheres \citep[e.g.][]{Owen2017,Wu2019,Gupta2019}. The exact extent to which these interiors were consistent with Earth was statistically quantified in \citet{Rogers2021,Rogers2023}, who found that such interiors were, in fact, $\sim 10\%$ under-dense when compared to Earth's density of $\sim 5.5 \text{ g cm}^{-3}$. This also aligns with the precise density measurements of the TRAPPIST-1 system planets, which demonstrate similar bulk under-densities when compared to Earth \citep{Agol2021}. 

Hydrogen is a highly reactive species, meaning chemical reactions with a magma ocean and core are heavily favoured. \citet{Schlichting2022} used chemical equilibrium models for sub-Neptunes, comprised of metal-rich cores, silicate-rich mantles and hydrogen-rich atmospheres, to demonstrate the overall reduction of cores and mantles, and oxidation of atmospheres. They noted that hydrogen and oxygen comprised significant fractions of the metal cores at chemical equilibrium, reducing their densities. \citet{Young2023} used the same chemical equilibrium models to argue that Earth's water content, core density deficit and overall oxidation state can be explained by the presence of an initial hydrogen-dominated atmosphere atop progenitor Earth embryos.

In this paper, we focus on super-Earths: those planets that are stripped of their hydrogen-dominated atmosphere in the scenario of atmospheric escape, leaving behind rocky bodies with negligible atmospheric mass, sitting below the radius gap. Despite losing their hydrogen-dominated atmospheres, we show that this chemical species will have left its mark on the planetary interior. The question is, then, can the prevalence of hydrogen in the history of super-Earths explain their reported under-densities when compared to Earth? Furthermore, what are the chemical properties and abundances of the escaping atmospheres, sculpted by mass loss and chemical equilibration?

\section{Method}\label{sec:Method}
We aim to determine the global chemical equilibrium state of super-Earths, under the assumption that they originally hosted a significant hydrogen-dominated atmosphere atop an Earth-like interior \citep[$\sim 1/3$ metal core, $\sim 2/3$ silicate mantle, by mass, as seems likely given recent surveys of probable metal core fractions e.g.][]{Trierweiler2023}. To do so, we utilise the chemical equilibrium model of \citet{Schlichting2022,Young2023}. Super-Earths likely form magma oceans in contact with their hydrogen-dominated primordial atmospheres during their early evolution. During this molten phase of evolution, one can assume that chemical equilibrium would have been maintained between the interior and atmosphere due to vigorous mixing and convection. However, the computational cost of our global chemical equilibrium model demands that we cannot feasibly track the state of chemical equilibrium through time. Therefore, we instead evaluate at the \textit{time of last global chemical equilibrium}, which we define as the point in time at which the atmosphere is no longer able to remain in chemical equilibrium with the interior due to substantial crystallisation of the magma ocean. Note that finding the time of last global chemical equilibrium is only possible for super-Earths, since they rapidly lose their hydrogen-dominated atmospheres, allowing the magma ocean to cool more and more efficiently through a diminishing atmosphere. On the other hand, the magma oceans of sub-Neptunes are likely to remain fully molten due to their large hydrogen-dominated atmospheres, preventing sufficient cooling and termination of global chemical equilibrium. To identify the time of last global chemical equilibrium for super-Earths, we use the semi-analytic atmospheric structure and evolution models of \citet{Rogers2021}, which we outline below.

\subsection{Determining the time of last global chemical equilibrium} \label{sec:LastGlobalChemEq}
To find the time of last global chemical equilibrium, we model the atmospheric evolution of a typical super-Earth. We assume a radiative-convective equilibrium model, following the works of \citet{Ginzburg2016,Owen2017,Gupta2019,OwenCamposEstrada2020,Rogers2021}. These models assume an interior convective region, modelled as an adiabat of index $\gamma$ with a temperature profile \citep[e.g.][]{Rafikov2006}:
\begin{equation} \label{eq:temperature_profile_adiabat}
    T(r) = T_\text{rcb}  \bigg[ 1 + \nabla_\text{ad} \frac{G M_\text{p}}{c_\text{s}^2 R_\text{rcb}} \bigg( \frac{R_\text{rcb}}{r} - 1 \bigg) \bigg],
\end{equation}
where the subscript `rcb' refers to quantities evaluated at the radiative-convective boundary. Here,  $\nabla_\text{ad} \equiv (\gamma - 1) / \gamma$ is the adiabatic gradient and $M_\text{p}$ is the planet's mass. Above the convective region sits a radiative layer, assumed to be isothermal at an equilibrium temperature of $T_\text{eq}$. {\jr{Note that, by definition of an isothermal outer region, $T_\text{eq} = T_\text{rcb}$.}} The sound speed in Equation \ref{eq:temperature_profile_adiabat} is evaluated at the radiative-convective boundary, $c_\text{s}^2 \equiv k_\text{B} T_\text{eq} / \mu m_\text{H}$, where $\mu=2.35$ is the mean molecular weight for an initially hydrogen-dominated atmosphere \citep[e.g.][]{Anders1989}. {\jr{Note that in Section \ref{sec:abundances}, we find that the atmospheric mean-molecular weight will increase by a factor of $\sim 2-4$ with time due to the effects of chemical equilibrium and atmospheric escape. If the atmosphere is fully mixed all the way to the sonic point, increasing the mean-molecular weight will, to first order, reduce mass loss rates since the sound speed of escaping gas reduces. For now, we assert that this change has little effect on the results of this study, and discuss this in detail in Section \ref{sec:Assumptions}. We do consider the escape of heavier species e.g. H$_2$O, within atmospheres in Section \ref{sec:MR_atmospheres}.}}

We evolve models from an initial atmospheric mass fraction, defined as $X_\text{atm} \equiv M_\text{atm} / M_\text{interior}$, where $M_\text{atm}$ is the atmospheric mass, and $M_\text{interior}$ is the non-gaseous planet mass (typically referred to as the `core' mass in exoplanetary science). We adopt the following initial atmospheric mass fractions:
\begin{equation} \label{eq:Ginzburg2016}
    X_\text{atm, init} = 0.01 \, \bigg( \frac{M_\text{interior}}{M_\oplus} \bigg)^{0.44}  \bigg( \frac{T_\text{eq}}{1000 \text{ K}} \bigg)^{0.25},
\end{equation}
from the analysis of \citet{Ginzburg2016}, which accounts for initial gas accretion and boil-off during protoplanetary disc dispersal {\jr{(assuming no orbital migration)}}. The model is evolved, accounting for radiative cooling and atmospheric escape \citep[see][for details]{Rogers2021}. For the latter process, we assume XUV photoevaporative mass-loss rates of:
\begin{equation} \label{eq:energy-limited-Mdot}
     \dot{M} = \eta \, \frac{\pi R_\text{p}^3 F_\text{XUV}}{G M_\text{p}},
 \end{equation} 
 where $F_\text{XUV}$ is the incident stellar XUV flux \citep[following][]{Rogers2021b}, $R_\text{p}$ is the planet's radius, and $\eta$ is the mass-loss efficiency, taken from the hydrodynamic models of \citet{OwenJackson2012}. For simplicity, we model atmospheric mass loss by photoevaporation rather than core-powered mass-loss \citep[e.g.][]{Ginzburg2018,Gupta2019}, but we stress that our results are independent of which of these two mass-loss mechanisms is assumed, given the similarities in the underlying physics \citep[e.g. see Figure 1 of][]{Rogers2023b}. In this study, we only concern ourselves with the atmospheric masses and interior temperatures at a single point in a planet's evolution, i.e. the time of last global chemical equilibrium in a planet's evolution, which are very similar for the two mass loss mechanisms.

 To determine the atmospheric conditions at the time of last global chemical equilibrium, we track the temperature at the base of the atmosphere, $T_\text{atm,base}$, {\jr{defined as $T(r=R_\text{interior})$ from Equation \ref{eq:temperature_profile_adiabat}}}. {\jr{We begin our evolution models at $10$~Myr, for which this base temperature can range from $\sim 4000-10,000$~K, depending on $M_\text{interior}$ and $X_\text{atm,init}$. As the planet thermally cools and loses mass due to atmospheric escape, $T_\text{atm,base}$ reduces. We define the time of last global chemical equilibrium as the time at which the temperature at the base of the atmosphere drops $T_\text{atm,base} \leq 2000$~K, appropriate for substantial crystallisation throughout a silicate melt \citep[e.g.][]{Andrault2011}. Such crystallisation is likely to halt large-scale mixing, and thus prevent efficient chemical equilibrium throughout the planet.}} At this point, we record the atmospheric mass fraction, $X_\text{atm,eq}$. Then, the goal is to find a global chemical equilibrium model that results in a hydrogen atmospheric mass fraction equal to this value $X_\text{atm,eq}$. In other words, we assert that the global chemical equilibrium state of a planet with a hydrogen atmospheric mass fraction of $X_\text{atm,eq}$ is indicative of a super-Earth, bound to lose the remaining hydrogen content of its atmosphere, and for which large-scale chemical reactions between interior and atmosphere are unlikely to continue. 

\vspace{1cm}
\subsection{Evaluating global chemical equilibrium} \label{sec:chemicalEq_solver}
To find global chemical equilibrium, we adopt the models of \citet{Schlichting2022,Young2023}. These models use a set of linearly independent chemical reactions, including 13 phase components for the liquid mantle and core: MgO, SiO$_2$, MgSiO$_3$ , FeO, FeSiO$_3$, Na$_2$O, Na$_2$SiO$_3$, Fe$_\text{metal}$, Si$_\text{metal}$, O$_\text{metal}$, H$_\text{metal}$, H$_\text{2,silicate}$, and H$_2$O$_\text{silicate}$; and 6 phase components for the atmosphere: H$_\text{2,gas}$, CO$_\text{gas}$, CO$_\text{2,gas}$, CH$_\text{4,gas}$, O$_\text{2,gas}$, and H$_\text{2}$O$_\text{gas}$. Reactions are allowed to take place between core and mantle, as well as between mantle and atmosphere. We list the linearly independent reactions in Appendix \ref{app:reactions}, although we stress that these reactions are not the only allowed reaction pathways. Instead, they are a set of basis vectors through which other relevant reactions may occur through linear combinations of such basis vectors \citep[see][for details]{Schlichting2022}.

 Chemical equilibrium is found by solving for the mole fraction of each species, $x_i$, such that for each reaction:
\begin{equation} \label{eq:ChemEq}
    \sum_i \nu_i \, \mu_i = 0,
\end{equation}
where the index $i$ is the label for each species in a reaction, $\nu_i$ is its stoichiometric coefficient, and $\mu_i$ is its chemical potential, given by:
\begin{equation} \label{eq:chemical_pot}
    \mu_i = \Delta \hat{G}_i^o + RT \ln (x_i),
\end{equation}
where $\Delta \hat{G}_i^o$ is the Gibbs free energy of formation at the standard state of the pure species $i$, corrected for pressure in the case of vapour species. Here, $R$ is the gas constant and $T$ is the temperature at which the reaction takes place. As discussed in detail in \citet{Schlichting2022}, Equation \ref{eq:ChemEq} is solved for the mole fraction of each species, along with the atmospheric pressure, all with the additional constraints of conservation of mass for all elements, and a unity sum of mole fractions for all species in each phase. Numerically, we solve the system of non-linear equations with the simulated annealing algorithm \verb|dual_annealing| of \citet{Xiang1997}, and a Monte Carlo Markov chain (MCMC), specifically the \verb|emcee| Python implementation \citep{ForemanMackey2014} of the affine-invariant ensemble sampler from \citet{GoodmanWeare2010}. Information on our adopted standard-state molar Gibbs free energies of reaction, $\Delta \hat{G}_i^o$, can be found in the Appendix of \citet{Schlichting2022}.

\begin{figure*}
	\includegraphics[width=2.0\columnwidth]{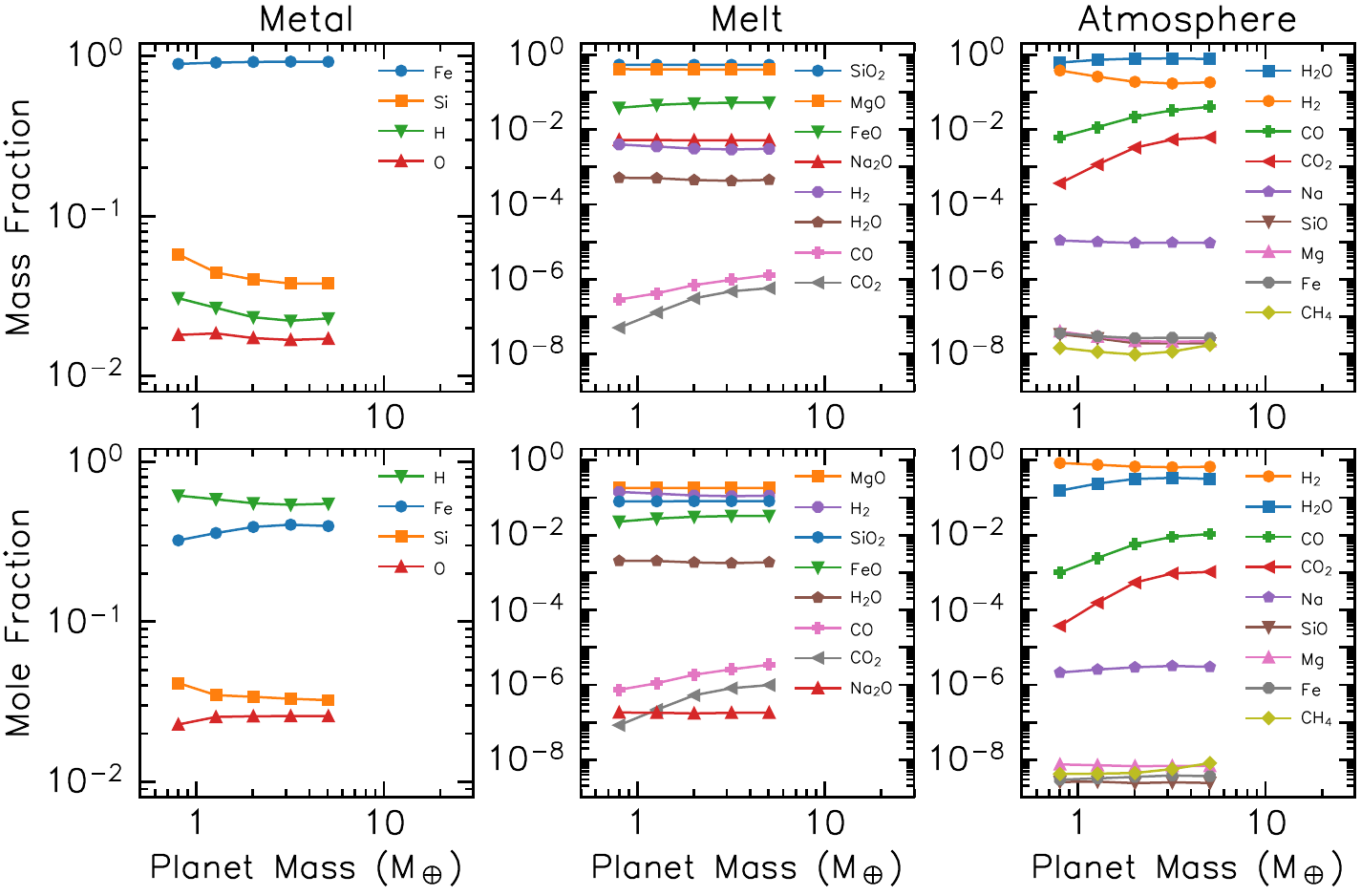}
    \centering
        \cprotect\caption{Chemical abundances of modelled super-Earths at the time of last global chemical equilibrium, in terms of mass fraction and mole fraction in the top and bottom row respectively. We show these abundances in the metal, silicate melt, and atmospheric phases for each species in the left, middle and right-hand columns respectively. We predict large quantities of light element (H and O) sequestration into the metal and the production of significant quantities of H$_2$O in the atmosphere.} \label{fig:combined_abundances_3000K} 
\end{figure*}

We set the reaction temperature, $T$ in Equation \ref{eq:chemical_pot}, to $2000$~K for reactions between atmosphere and silicate melt, such that the magma ocean is substantially crystallised \citep[e.g.][]{Andrault2011} thus preventing global chemical equilibrium via efficient mixing (see Section \ref{sec:LastGlobalChemEq}). Reactions between silicate melt and metal melt, on the other hand, are likely to occur at greater depths and, thus, higher temperatures. Therefore, we adopt a reaction temperature of $3000$~K between silicate and metal species, as is suggested to be the case for Earth's silicate-metal equilibration \citep[e.g.][]{Wood2008}. The simplicity in these choices can be justified in this very first study, given the uncertainty in the interior structure of rocky super-Earths. An example of such uncertainties is that, as is the case for Earth, one may expect metal to rain out from the silicate to form an iron-rich core due to its higher density. In this case, an adiabatic temperature profile can be used to calculate the reaction temperature between the metal and silicate species (at the core-mantle boundary, CMB). However, for the early planetary formation of super-Earths in hydrogen-rich nebulae, it remains unclear whether this core differentiation does, in fact, occur \citep[e.g.][]{Lichtenberg2021}. As shown in \citet{Schlichting2022,Young2023}, and as we shall demonstrate in Section \ref{sec:abundances}, hydrogen is sequestered extremely efficiently inside the interior. Its effect will be to alter the buoyancy of both metal and silicate species, potentially changing convection, equation of state properties, and the ultimate rain-out of dense metal species. Choosing the correct temperature profile is, therefore, fraught with many of these uncertainties, which we discuss further in Section \ref{sec:Assumptions}.

As discussed in Section \ref{sec:LastGlobalChemEq}, the goal is to find a global chemical equilibrium model for which the hydrogen mass fraction in the atmosphere is equal to that which renders a magma ocean surface temperature of $2000$~K, as a result of atmospheric cooling and mass loss. We refer to this hydrogen atmospheric mass fraction as $X_\text{atm,eq}$. We stress again that this is only possible for super-Earths, and not for sub-Neptunes, since the latter will likely retain fully molten magma oceans due to their thick hydrogen-dominated atmospheres. In our case of super-Earths, we run successive equilibrium models with increasing amounts of hydrogen already sequestered in the interior (mostly in the form of H$_\text{metal}$), as well as an initial hydrogen atmospheric mass fraction given from Equation \ref{eq:Ginzburg2016}, until the amount of hydrogen in its atmosphere matches $X_\text{atm,eq}$, with a numerical accuracy of $<2\%$. The accepted model is, therefore, representative of a rocky super-Earth, having formed in a hydrogen-rich protoplanetary disc, subjected to atmospheric cooling and escape, until its interior no longer chemically interacts with its atmosphere.

\section{Results} \label{sec:Results}
We model super-Earths with an atmospheric equilibrium temperature of $1000$~K {\jr{(assuming $0$ albedo)}} and masses between $0.8M_\oplus \leq M_\text{p} \leq 5.1M_\oplus$, which encompass the range of planet masses that would have been stripped via atmospheric escape at this equilibrium temperature around a solar mass star. We assume an Earth-like interior iron mass fraction of $0.32$ for all planets and a hydrogen-dominated atmosphere with only trace amounts of other gaseous species. We now present results on their bulk properties and chemical makeup at the time of last global chemical equilibrium in the following sections.

\subsection{Interior and atmospheric chemical compositions} \label{sec:abundances}

\begin{figure*}
	\includegraphics[width=1.8\columnwidth]{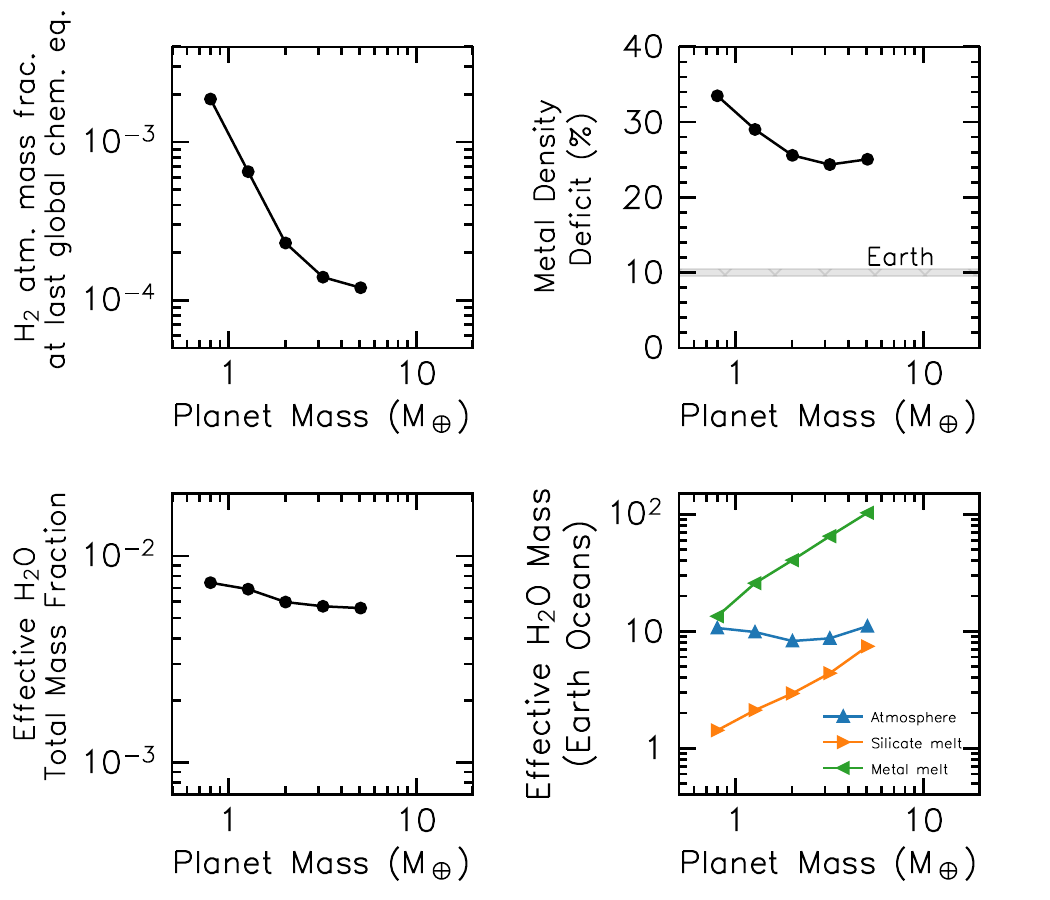}
    \centering
        \cprotect\caption{Upper left: the H$_2$ atmospheric mass fraction at the time of last global chemical equilibrium, defined as the time a super-Earth's magma ocean substantially crystallises, thus halting large-scale chemical reactions between interior and atmosphere. Upper right: the metal density deficit, relative to pure Fe, is shown for varying planet masses. For reference, this is compared to that of Earth, which has a measured metal deficit of $\sim 10\%$ \citep[e.g.][]{Birch1964,Badro2015}. Lower left: the total effective water mass fraction represents the amount of water endogenously produced due to chemical equilibrium with a hydrogen atmosphere. Lower right: the effective water mass reservoirs, measured in Earth oceans ($1.4 \times 10^{24}$~g), split into individual gaseous, silicate melt and metal phases in blue orange and green respectively. Here, we consider \textit{effective} content since molecular H$_2$O is very unlikely to exist, specifically in the metal core. In practice, the effective water content is calculated by pairing all O atoms in the metal phase with two H atoms. Note that, as shown in Figure \ref{fig:combined_abundances_3000K}, this pairing is oxygen-limited.} \label{fig:CombinedProperties} 
\end{figure*} 

We calculate the mole and mass fractions at chemical equilibrium of each species, in each phase, which are presented in Figure \ref{fig:combined_abundances_3000K}. Abundances in the metal, silicate melt, and atmosphere are shown in the left, middle and right-hand panels, respectively. Mass and mole fractions are shown in the top and bottom rows, respectively. 

\subsubsection{Interior} \label{sec:results_interior}
By mass, iron constitutes $\gtrsim 90\%$ of the metal across the entire range in planet masses. However, in terms of mole fraction, there is approximately twice as much hydrogen than iron, the former having been sequestered into the metal as a direct result of equilibration with a hydrogen-rich atmosphere. There is also $5$~\% Si and $2$~\% O, by mass, as a result of reduction/oxidation reactions triggered by interactions with hydrogen. The presence of light elements in the metal leads to a metal density deficit when compared to pure iron, which is shown in the upper right-hand panel of Figure \ref{fig:CombinedProperties} \citep[see][for details of this calculation]{Schlichting2022,Young2023}. One can see that larger-mass planets experience a lower metal density deficit. This is because larger-mass planets host atmospheres with steeper temperature gradients in their convective envelopes, leading to less atmospheric mass being present at the time of last global chemical equilibrium (as defined in Section \ref{sec:LastGlobalChemEq}). As a result, larger-mass planets undergo the time of last global chemical equilibrium with a smaller hydrogen mass fraction, ultimately acting as a smaller reservoir for any metal density deficit, as shown in the upper-left panel of Figure \ref{fig:CombinedProperties}. For reference, Earth has a measured metal density deficit of $10\%$ \citep[e.g.][]{Birch1964,Badro2015}. We predict a larger deficit than Earth, which is unsurprising since Earth's final assembly likely proceeded from roughly Mars-sized embryos via a giant impact phase long after the gas disc had dissipated. Intriguingly, it was recently shown that repeating the above global chemical equilibrium calculations for Earth's progenitor embryos, does yield a density deficit consistent with Earth \citep{Young2023}.

The silicate melt mass fraction is dominated by the metal oxides SiO$_2$, MgO and FeO. This is unsurprising since these species are the constituents of typical silicates such as MgSiO$_3$, FeSiO$_3$ and Na$_2$SiO$_3$. Hydrogen is present in the silicate, but does not contribute a significant density deficit, unlike the metal, with mass fractions of $\lesssim 0.5\%$ for H$_2$ and $\lesssim 0.05\%$ for H$_2$O.    

\subsubsection{Atmosphere} \label{sec:results_atm}
The atmospheres of super-Earths at the time of last global chemical equilibrium can be thought of as transient atmospheres. As an initial sub-Neptune loses its hydrogen-dominated atmosphere, physical processes will continue to sculpt the atmosphere's composition as it transitions to become a super-Earth with a negligible atmospheric mass. Here, we present the chemical state of super-Earth atmospheres at the time of last global chemical equilibrium, which corresponds to $\sim$ a few tens of Myrs. It is important to note that the final state of these atmospheres after $\sim$~Gyrs of evolution will continue to be dictated by other processes, such as further atmospheric escape or outgassing, which we discuss in Sections \ref{sec:MR_atmospheres} and \ref{sec:Discussion}.

As shown in Figure \ref{fig:combined_abundances_3000K}, we find atmospheres that are steam-dominated by mass fraction at the time of last global chemical equilibrium. This corresponds to hydrogen-dominated atmospheres by mole (number) fraction.  Although we do not solve for chemical abundances as a function of pressure and temperature in the atmosphere \citep[e.g.][]{Markham2022,Misener2022,Misener2023}, we do calculate mean-molecular weights,\footnote{To calculate the atmospheric mean-molecular weight, we only consider H$_2$, H$_2$O, CO, CO$_2$ and CH$_4$ and O$_2$ species, since the remaining gas species would condense out and not be observable in the upper atmosphere.} which range from $4.5 \lesssim \mu / m_\text{H} \lesssim 7.4$. {\jr{The main driver of the increased mean molecular weight is the production of water, which constitutes a partial pressure in the range $\sim 40 - 400$~bar for the planets in this study.}} This water reservoir is unlikely to condense out into an ocean phase since these planets exist at such extreme equilibrium temperatures, thus precluding the ability for the steam to cool. Apart from the residual H$_2$, the most abundant atmospheric species by mole fraction after H$_2$O are CO and CO$_2$, present at the $\sim0.01-1\%$ level by atmospheric mass, with CO $/$ CO$_2$ number ratios of $\sim 20$. We note here that this relative ratio of  CO $/$ CO$_2$ is only representative in the deep atmosphere as photochemistry is expected to alter this ratio in the upper atmosphere, which can be probed observationally. Trace amounts of the remaining gas constituents, such as SiO and CH$_4$, are also present. Note that O$_2$ is not included in Figure \ref{fig:combined_abundances_3000K}, due to its very low abundance of $\sim 10^{-11}$ by mole fraction. As discussed, we do not consider the atmospheric structure and resultant variation in chemical abundances as a function of pressure and temperature above the planet's surface, as done in \citet{Markham2022,Misener2022,Misener2023}. This added complexity is beyond the scope of this work and is left for future study. 

\subsubsection{Global water content} \label{sec:results_H2Obudget}
As discussed, Figure \ref{fig:CombinedProperties} shows the H$_2$ atmospheric mass fraction at the time of last global chemical equilibrium, as well as the metal density deficit in the Fe core in the top-left and top-right hand panels respectively. In addition, we also show the global water content of super-Earths, both as a total mass fraction and also split between metal, silicate melt and atmospheric components in the bottom-left and bottom-right hand panels respectively. Note that, although it is commonplace to discuss the interior water content of planets \citep[e.g.][]{Dorn2021}, H$_2$O is very unlikely to exist in molecular form within the deep interior, particularly in the iron core. Instead, H and O will exist in atomic form (hence our speciation in Figure \ref{fig:combined_abundances_3000K}). Nevertheless, we compute the total \textit{effective} water content of the planet in each phase for comparison. In the case of the metal core, this is performed by pairing oxygen atoms with two hydrogen atoms until the oxygen reservoir is depleted. One can see that super-Earths host $\sim 0.5\% - 1.0\%$ H$_2$O by mass as a direct result of chemical equilibrium with a hydrogen-dominated atmosphere. We refer to this as \textit{endogenously} produced water. Most of this effective water content is stored in the metal core, with approximately an order of magnitude less in the mantle, as shown in the bottom right-hand panel of Figure \ref{fig:CombinedProperties}. The atmosphere, on the other hand, stores an intermediate amount of water in the form of steam (although this is likely removed during further atmospheric escape, see Section \ref{sec:MR_atmospheres}). The atmosphere holds $\sim 10$ Earth oceans of H$_2$O, which is independent of planet mass due to the approximate balance of competing effects: on one hand, larger planets host a larger reservoir of chemical reactants to produce water; on the other hand, larger super-Earths also retain smaller H$_2$ atmospheres at the time of last global chemical equilibrium, as shown in the top-left hand panel of Figure \ref{fig:CombinedProperties}.

\begin{figure}
	\includegraphics[width=1.0\columnwidth]{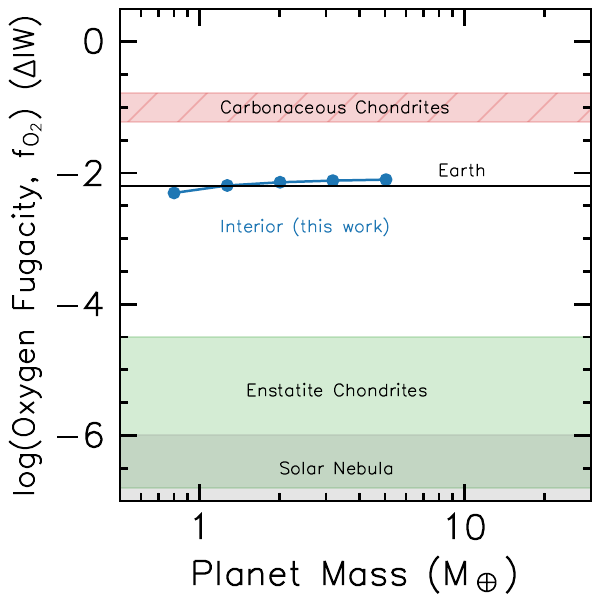}
    \centering
        \cprotect\caption{The oxygen fugacity of the silicate melt is shown for varying super-Earth masses in blue. Fugacity is measured relative to the FeO (Iron Wüstite, IW) buffer. We compare these to the values for Earth \citep{Doyle2019}, as well as carbonaceous chondrites (red-shaded region), enstatite (E) chondrites (green-shaded region) and the Solar nebula (grey-shaded region).} \label{fig:fugacity} 
\end{figure}

\subsubsection{Oxygen fugacity} \label{sec:results_fO2}
In addition to chemical abundances, we also compute the oxygen fugacity $f_{\text{O}_2}$ of the magma ocean and atmosphere. Fugacity measures the partial pressure of oxygen that would be in equilibrium with the system and is, therefore, a convenient measure of a material's oxidation state at equilibrium. Historically, this is measured relative to the reaction in which pure iron is oxidised to form pure FeO (iron wüstite, IW), Fe + $1/2$ O$_2$ = FeO. When reported as deviations from the IW reference, on a logarithmic scale, the oxygen fugacity is reported as:
\begin{equation}
     \Delta \text{IW} = 2\log \bigg (\frac{x_\text{FeO}}{x_\text{Fe}} \bigg),
\end{equation}
where $x_i$ refers to a mole fraction of FeO in the silicate melt or Fe in the metal. As with the issue of \textit{effective} molecular H$_2$O in the metal core (as described above), molecular H$_2$ is very unlikely to exist in the melt. Instead, atomic hydrogen will exist in the silicate crystals and oxides in polymer form. Therefore, the fugacity of the magma ocean is a measure of the partial pressure of O$_2$, if it indeed existed. Following this convention, we find that the fugacities of our super-Earth's magma oceans are remarkably similar to that of Earth, which has a value of $\Delta \text{IW} \approx -2.2$ \citep{Doyle2019}, as shown in Figure \ref{fig:fugacity}. This was also found in \citet{Young2023} when considering the chemical equilibrium of a proto-Earth and is explained due to the oxidation of Fe displaced from the core by Si. Overall, we see an increase in the oxidation state of the magma ocean from that of the solar-like nebula and E chondrites to that of Earth, purely due to the chemical equilibrium of a magma ocean with a hydrogen atmosphere. Despite the apparent contradiction of hydrogen (a reductant) causing the net oxidation of the interior, this is, in fact, due to the reduction of Si, causing a balancing oxidation of Fe \citep{Schlichting2022,Young2023}.

\subsection{Mass-radius relations and bulk densities} \label{sec:massradius}
In this Section, we compare our models with observations of super-Earths. We begin with a prescription for determining mass-radius relations, followed by an outline of our data selection and analysis.

\begin{figure*}
	\includegraphics[width=2.0\columnwidth]{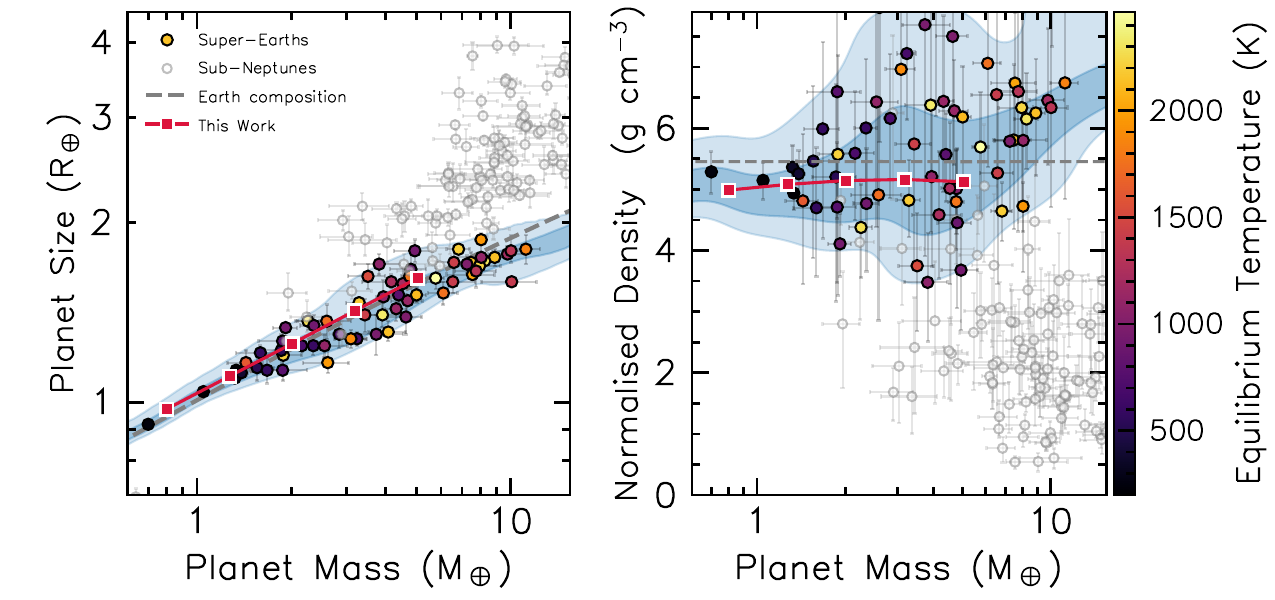}
    \centering
        \cprotect\caption{{\jr{Our super-Earth models are shown in pink squares, compared with a sample of observed exoplanets from the \verb|PlanetS| catalogue \citep{Otegi2020} with masses and radii constrained to $25\%$ and $8\%$ respectively. In the left-hand panel, we show the mass-radius diagram, whereas, in the right-hand panel, we show normalised density as a function of planet mass, which is the density of each planet when scaled to be $1 M_\oplus$. Grey dashed lines represent an Earth-like density of $\sim 5.5 \text{ g cm}^{-1}$. The blue-shaded regions represent the $1\sigma$ and $2\sigma$ ranges in radius and density for the population of super-Earths, as laid out in Section \ref{sec:DataSelection}.}}} \label{fig:MR} 
\end{figure*}

\subsubsection{Determining Mass-Radius Relations} \label{sec:MR_atmospheres}
We calculate planetary radii by following \citet{Rogers2010} and solving for the radius $r(m)$ and pressure $P(m)$ of a spherically symmetric body in hydrostatic equilibrium as a function of interior mass, $m$:
\begin{equation} \label{eq:HSE1}
    \frac{\partial r}{\partial m} = \frac{1}{4 \pi r^2 \rho},
\end{equation}
\begin{equation} \label{eq:HSE2}
    \frac{\partial P}{\partial m} = -\frac{Gm}{4 \pi r^4},
\end{equation}
where $\rho$ is the density, given by an equation of state:
\begin{equation}
    \rho = f(P, T),
\end{equation}
which relates the density to the pressure $P$ and temperature $T$ within each layer. For each planet of mass $M_\text{p}$, we integrate Equations \ref{eq:HSE1} and \ref{eq:HSE2} from $0 \leq m \leq M_\text{p}$ and impose inner boundary conditions of $r(m=0) = 0$ and $P(m=0)=P_\text{c}$, where $P_\text{c}$ is the central pressure. We solve for $P_\text{c}$, which yields a planet size, such that $P(m=M_\text{p})=P_\text{atm}$. We integrate Equations \ref{eq:HSE1} and \ref{eq:HSE2} with \verb|scipy.integrate.solve_ivp| and solve for the planet radius with \verb|scipy.optimize.newton| \citep{SciPy2020} with fractional tolerances of $10^{-8}$. 

We assume that, after Gyrs of evolution, planets will have formed differentiated iron-rich cores (an assumption that we question in Section \ref{sec:CoreDiff}). As such, we use a temperature-independent equation of state for Fe from \citet{Schlichting2022}, following \citet{Seager2007,Anderson2001}. For the silicate melt, we use the Mg-perovskite third-order Birch-Murnagham equation of state of \citet{Karki2000}, as described in \citet{Seager2007}. We assume an Earth-like iron-core mass fraction of $0.32$, incorporating a metal density deficit relative to pure iron. This arises from the sequestration of light species such as hydrogen and oxygen, reducing the overall density of the metal core.  We determine the metal density deficit for each model (as shown in Figure \ref{fig:CombinedProperties}) and calculate the planet's radius at the base of the atmosphere. 

The presence of an atmosphere, if sufficiently optically thick, will also contribute to a planet's size. Indeed, the atmospheric masses of our modelled super-Earths at the time of last global chemical equilibrium (occurring at $\sim$ a few tens of Myrs) are sufficient to produce a small increase in planet size. However, the atmosphere of a super-Earth observed after Gyrs of evolution is expected to be smaller in mass than those at the time of last global chemical equilibrium. This is because atmospheric escape, driven through photoevaporative and/or bolometrically driven outflows, will continue to erode the atmosphere. Specifically, escaping hydrogen, which is the lightest and easiest to liberate from the planet, can drag heavier species along with it. One can justify this with a simple calculation of the cross-over mass $m_\text{c}$, which represents an estimate of the maximum atomic/molecular mass species that can be dragged along within the hydrogen outflow \citep[e.g.][]{Hunten1987,Chassefiere1996,Luger2015}:
\begin{equation}
    m_\text{c} = m_{\text{H}_2} + \frac{k_\text{B} T}{4 \pi b \, m_{\text{H}_2} G M_\text{p}} \dot{M}_{\text{H}_2}.
\end{equation}
Here, $m_{\text{H}_2}$ is the molecular mass of H$_2$, $T$ is the temperature of the outflow,\footnote{The outflow temperature can range from approximately the equilibrium temperature $T_\text{eq}$ in a bolometrically heated outflow, to $\sim 10^4$~K in a photoevaporative outflow \citep[e.g.][]{MurrayClay2009,Schulik2022,OwenSchlichting2023}.} $b \approx 4.8 \times 10^{17} \; (T / \text{K})^{0.75} \text{ cm}^{-1} \text{ s}^{-1}$ is the binary diffusion coefficient for the two species \citep{Zahnle1986}, and $\dot{M}_{\text{H}_2}$ is the hydrogen mass loss rate. For the super-Earths considered in this study, and in the simpler case of a bolometrically heated, neutral, isothermal outflow {\jr{for which we set the outflow temperature equal to $T_\text{eq}$ \citep[as is the case for core-powered mass loss, e.g.][]{OwenSchlichting2023}, mass loss rates are approximated using a Parker wind model:
\begin{equation} \label{eq:parker_mlr}
    \dot{M}_{\text{H}_2} = 4 \pi \, R_\text{B}^2 \, c_\text{s} \, \rho_\text{rcb} \, \exp \bigg\{ \frac{3}{2} - \frac{2R_\text{B}}{R_\text{rcb}} \bigg\},
\end{equation}
where $R_\text{B} \equiv G M_\text{interior} / 2 c_\text{s}^2$ is the Bondi radius. For the super-Earths considered in this study, Equation \ref{eq:parker_mlr} yields typical hydrogen mass loss rates of $\dot{M}_\text{H} \sim 10^{11} \text{ g s}^{-1}$ at the time of last global chemical equilibrium \citep[see Figure $6$ from][]{Misener2021}.}} In these cases, one attains cross-over masses of order $m_\text{c} \sim 10^2 - 10^4 m_\text{H}$, which are significantly larger in mass than any gas species within the atmosphere. We use the semi-analytic cross-over mass prescription from \citet{Luger2015} to estimate the evolution of each super-Earth modelled in this paper, including cross-over mass and escape rates for H$_2$ and H$_2$O. In case of a bolometrically heated, neutral, isothermal outflow, we confirm that the final atmospheric mass fractions of our super-Earths are $\lesssim 10^{-5}$, rendering a sufficiently small {\jr{atmospheric optical depth}} so as not to increase the planet's size. Whilst these simple calculations are informative, they ignore effects such as XUV ionisation and the complex cooling processes of molecular species in optically thin outflows. In an ionised outflow, higher mean-molecular weight species have an increased collisional cross-section \citep[e.g.][]{Geiss1970,Joselyn1978,Geiss1982}, which allows such species to be dragged more efficiently within the outflow. This suggests that heavier species are likely lost at a higher rate than estimated using the cross-over mass calculations discussed above once the contribution from ionising radiation is considered self-consistently. Understanding these processes in more detail requires hydrodynamic modelling in the context of multiple chemical and ionised species under high-energy irradiation environments, which will then allow for robust predictions of the final state of super-Earth atmospheres, which is beyond the scope of this work.

\subsubsection{Data Selection and Comparison} \label{sec:DataSelection}
{\jr{We compare our models with a high-fidelity sample of super-Earths with reliably measured masses and radii. We begin with the \verb|PlanetS| catalogue \citep{Otegi2020}, which selects planets based on various reliability and precision conditions. Planets in this catalogue have relative mass and radius uncertainties less than $25\%$ and $8\%$, respectively. From there, we define a super-Earth to exist below the radius valley. Since the exact definition of the radius valley varies within the literature, we take a conservative approach. From \cite{Ho2024}, in which short-cadence \textit{Kepler} photometry was used to produce a high-precision sample of transiting exoplanets $\lesssim 4R_\oplus$, the valley is inferred to follow:
\begin{equation} \label{eq:valley}
    \log \bigg(\frac{R_\text{p}}{R_\oplus} \bigg) = A\log \bigg(\frac{P}{1\text{ day}} \bigg) + B\log \bigg(\frac{M_*}{M_\odot} \bigg) + C,
\end{equation}
where $A=-0.09^{+0.02}_{-0.03}$, $B=0.21^{+0.06}_{-0.07}$, $C=0.35^{+0.02}_{-0.03}$. Note the importance of defining the valley as a function of orbital period $P$ and stellar mass $M_*$ due to its well-documented dependencies \citep[e.g.][]{Berger2020b,Gupta2020,Rogers2021b,Petigura2022}. We calculate the lower boundary of the valley by resampling Equation \ref{eq:valley} within its $1\sigma$ uncertainties $10^7$ times. The lower $1\sigma$ surface of these samples then defines the lower boundary as a function of orbital period and stellar mass. We define a super-Earth as a planet that sits below the lower boundary, thus removing planets potentially sitting within the radius valley, whose physical properties are currently unclear. The only exceptions to this classification scheme are TRAPPIST 1 planets, which are misclassified as sub-Neptunes under this scheme despite being terrestrial in nature \citep{Agol2021},\footnote{This misclassification arises from extrapolating the radius valley definition from Equation \ref{eq:valley} to very low stellar masses, such as that for TRAPPIST 1.} and GJ 367 b, which is consistent with a density of pure iron and hence considered an outlier \citep{Lam2021,Goffo2023}. As a result, we remove this planet from our sample.

We calculate each planet's \textit{normalised density} (often referred to as `compressed'), which is a planet's density if it had the same mass as Earth. In other words, planets that are over-dense when compared to Earth sit above $\sim 5.5 \text{ g cm}^{-3}$, and vice versa. For this comparison, we define an Earth-like composition to have a $32\%$ iron mass fraction with a constant $10\%$ metal density deficit \citep{Birch1964,Badro2015}. We then create a uniform grid in planet mass vs. normalised density space and use a Gaussian kernel density estimator to determine the probability density function of observed planets within this space. We normalise each grid slice, $i$, in planet mass, $M_\text{p,i}$, to contain equal probability density such that information is evenly spaced across the planet mass domain. Then, and again for each slice in planet mass $M_\text{p,i}$, we find the $1\sigma$ and $2\sigma$ contours for super-Earth normalised densities. These are presented in the right-hand panel of Figure \ref{fig:MR} as blue contours. The left-hand panel also contains the same contours converted to the mass-radius plane. Super-Earths, as classified in the discussion above, are shown as circles, coloured with their equilibrium temperature (assuming zero albedo). Sub-Neptunes are shown in grey. The $1\sigma$ and $2\sigma$ contours indicate a tentative under-density of super-Earths, which is consistent with the inference analysis of \citet{Rogers2021}. Our super-Earth models are shown in pink squares, which demonstrate an under-density when compared to Earth, due to hydrogen and oxygen sequestration into the metal core. The consistency with the data suggests that primordial hydrogen atmospheres interacting with young molten interiors provide a possible explanation for a density deficit in the super-Earth population. }}

\section{Discussion} \label{sec:Discussion}
This study has demonstrated the lasting effects of hydrogen on super-Earth interiors and atmospheres under the assumption that they initially hosted H$_2$-dominated atmospheres, which were then lost due to atmospheric escape. We now discuss the implications of these results and the possibility of observational tests.

\subsection{Do metal cores differentiate?} \label{sec:CoreDiff}
As additionally shown in \citet{Schlichting2022, Young2023}, we also find that hydrogen is efficiently sequestered into the interior of low-mass exoplanets in the presence of hydrogen atmospheres. Figure \ref{fig:combined_abundances_3000K}, in fact, demonstrates that hydrogen atoms outnumber iron by a factor of $\sim2$ within the metal. These results, including the water production as discussed in Section \ref{sec:abundances}, are in agreement with high $P-T$ experimental results from \citet{Horn2023,Piet2023}. From a geocentric view, one would expect iron species to rain out from the silicate melt over time to eventually form a metal core. However, the abundance of hydrogen may change this picture. Its presence, and willingness to bond with iron, may change the overall buoyancy within the melt, thus potentially delaying or even stopping metal rain-out in the interior. This was highlighted in \citet{Lichtenberg2021}, who showed that this `rain out quenching' is also controlled by the iron droplet size and the internal heat flux of the mantle. We stress that our chemical equilibrium calculations do not include an interior structure model. It is, therefore, beyond the scope of this work to self-consistently model the evolving rain out process in detail, including the solidification of the magma ocean below its surface. Nevertheless, the abundance of hydrogen may have other impacts on the global properties of the planet, such as changes in the efficiency of a potential dynamo mechanism and thus the prevalence of large-scale magnetic fields on super-Earths \citep[e.g.][]{J_Zhang2022}.

\subsection{Observational tests} \label{sec:discussion_obs}
It is well-documented that attempting to discern a planet's interior structure is fraught with degeneracies when only observing planetary masses and radii \citep[e.g.][]{Rogers2015,Bean2021,Rogers2023b}. Unsurprisingly, then, our population-level prediction of bulk under-densities, when compared to Earth (see Figure \ref{fig:MR}), does not constitute direct observational evidence of hydrogen sequestration. Additional mechanisms to produce bulk under-densities include a planet hosting a reduced iron-mass fraction \citep[as may be the case for older planets around metal-poor stars e.g.][]{Trierweiler2023}, or an increased initial volatile mass fraction. 

On the other hand, we also predict that super-Earths will enter a phase in which their atmospheres are steam-dominated by mass (although hydrogen-dominated by mole fraction) as a result of global chemical equilibration and atmospheric escape (see Section \ref{sec:results_atm}). {\jr{This is similar to the conclusions of \citet{Kite2021}.}} Perhaps the most interesting targets to detect such atmospheres are planets found inside the radius gap, since they may be in the process of transitioning from being a sub-Neptune to a super-Earth. In the era of atmospheric characterisation of small planets with observatories such as \textit{JWST}, it may be possible to detect such signatures in the coming years. We stress again, however, that further work is needed to understand how such atmospheres would evolve after the time of last global chemical equilibrium due to cooling, mass loss and outgassing. Furthermore, our chemical equilibrium models do not consider the variation of atmospheric chemical abundances with pressure and temperature, as was done in \citet{Markham2022,Misener2022,Misener2023}.

\subsection{Model assumptions and uncertainties} \label{sec:Assumptions}

In solving for global chemical equilibrium, we have made several simplifying assumptions in our model. One such assumption is that global chemical equilibrium is maintained within and between the interior and atmosphere via efficient bulk transport, convection and mixing until the time of last global chemical equilibrium \citep[e.g.][]{Lichtenberg2021,Salvador2023}. Although this is likely a reasonable first approximation, it would be desirable to relax this assumption in future work by coupling an evolving interior structure model \citep[e.g.][]{curry2024} with our chemistry model. This would allow one to track the cooling and crystallisation of silicates as a function of pressure and temperature, leading to the locking of volatiles inside the interior as a function of time.

As discussed in Section \ref{sec:LastGlobalChemEq}, we have assumed the silicate melt and atmosphere chemically equilibrate at $2000$~K, whereas the silicate melt and metal equilibrate at $3000$~K at the time of last global chemical equilibrium for all super-Earths. These temperatures are sensibly chosen to ensure significant magma ocean crystallisation such that efficient bulk transport and chemical equilibration are significantly suppressed beyond this point in time \citep[e.g.][]{Stixrude2014,Salvador2023}. {\jr{The silicate-atmosphere temperature of $2000$~K was chosen under the assumption that the magma ocean will solidify from the bottom up, in which case a surface temperature closer to the adiabatic silicate melt solidus $\sim 1500$~K would imply the vast majority of the mantle has already solidified and chemical equilibrium has already ceased. The higher value of $2000$~K thus captures the time at which the bulk of the magma ocean is in the process of crystallising. In a similar manner, we have chosen a silicate-metal reaction temperature of $3000$~K, which is lower than that predicted by an adiabatic silicate melt at the core-mantle boundary \citep[e.g.][]{Stixrude2014}. This is because we cannot assume the silicate melt and metal equilibrate at this temperature, especially when the planet is young and a differentiated core has not necessarily formed.}} This simplification, however, ignores the pressure dependence on the solidus/liquidus of an adiabatic silicate melt, which would introduce a planet mass dependence on these temperatures. As shown in \citet{Schlichting2022}, the temperature at which reactions take place can change the degree of light-element sequestration into the interior. {\jr{Although exploring the effects of varying reaction temperatures is beyond the scope of this work, one would expect higher temperatures to result in more light elements being sequestered into the metal core \citep[see Figure $12$ from ][]{Schlichting2022}.}} Sequestration is also affected by the adopted H$_2$O solubilities \citep[see][]{Schlichting2022}, which affect the partitioning of water into the interior \citep[e.g.][]{Moore1998,Dorn2021,Bower2022}. Nevertheless, we confirmed that the general result of light element sequestration and water production is robust to minor changes in the exact H$_2$O solubilities. As with the previous point, evolving structure models are required to self-consistently determine how efficiently the interior allows for global chemical equilibrium \citep[e.g.][]{curry2024} as a planet evolves and cools.

{\jr{Finally, in Section \ref{sec:LastGlobalChemEq}, we assumed that we can ignore, to first order, the increase in atmospheric mean-molecular weight from $\mu = 2.35$ to $4.5 \lesssim \mu / m_\text{H} \lesssim 7.4$, as found in Section \ref{sec:results_atm}, in our atmospheric escape models. As previously discussed, if the atmosphere is fully mixed all the way up to the sonic point, then increasing the mean-molecular weight reduces mass loss rates due to a decreased sound speed. However, since we are only concerned with the conditions of last global chemical equilibrium in this study, this change is unimportant in setting the hydrogen mass fraction, $X_\text{atm,eq}$, that renders a magma ocean surface temperature of $2000$~K, as considered here. In other words, changing mass loss rates may alter the exact time at which last global chemical equilibrium occurs, but the value of $X_\text{atm,eq}$, which sets the conditions for chemical equilibrium (see Section \ref{sec:chemicalEq_solver}) remains approximately the same. As highlighted in Section \ref{sec:MR_atmospheres}, the details of atmospheric escape in the presence of heavier gas species are complex and warrant further detailed study since the level of mixing and the resulting mean-molecular weight at various heights in the atmosphere need to be modelled self-consistently with mass loss. We, therefore, leave coupled models of atmospheric escape and chemical equilibrium for future work.}}

\section{Conclusions} \label{sec:Conclusions}
The likely fate of super-Earths is to have their primordial hydrogen-dominated atmospheres removed via thermally-driven atmospheric escape \citep[e.g.][]{LopezFortney2013,Owen2013}. In this study, we consider the effects of such mass loss, combined with global chemical equilibration, in order to understand the bulk and chemical properties of super-Earths. To do so, we use the models of \citet{Schlichting2022,Young2023} to evaluate the chemical state of planets at the \textit{time of last global chemical equilibrium}, defined as the time at which large-scale chemical interactions cease to occur between an atmosphere and planetary interior due to significant crystallisation of a magma ocean. Our conclusions are as follows:

\begin{itemize}
    \item The net result of chemical equilibration is the efficient sequestration of hydrogen into the planetary interior. This produces bulk densities of $\sim 5.0 \text{ g cm}^{-3}$ when compared to Earth's density of $5.5 \text{ g cm}^{-3}$. This arises due to light elements, such as hydrogen and oxygen, found in the metal phase as they oxidise iron. This produces metal density deficits of $20\%-40\%$. We show that these bulk densities are consistent with the exoplanet population and evolution inference analysis from \citet{Rogers2021}.

    \item A by-product of hydrogen interactions with the interior is the production of significant quantities of water, up to $\sim 0.5 - 1$\% of a planet's mass. This equates to tens, to hundreds of Earth oceans being present within super-Earths, produced entirely through endogenous means. The water exists predominantly as a reservoir in the metal phase (in the form of separated oxygen and hydrogen), but also in the atmosphere and magma ocean to a lesser extent.

    \item As a super-Earth loses its hydrogen-dominated atmosphere, its remaining atmosphere will become steam-dominated by mass (although still hydrogen-dominated by mole fraction) as a result of chemical equilibration {\jr{\citep[see also][]{Kite2021}}}. We evaluate the chemical abundances of super-Earth atmospheres at the time of last global chemical equilibrium, which demonstrate mean-molecular weights of $4.5 \lesssim \mu / m_\text{H} \lesssim 7.4$. Carbon-bearing species such as CO and CO$_2$ are also expected to be present at the $\sim0.01-1\%$ by mass at this time. We stress that this atmosphere represents a single snapshot during a super-Earth's evolution and will inevitably be sculpted by further mass loss and outgassing. Nevertheless, these transient, steam-dominated atmospheres may be observable for planets transitioning across the radius valley with current observational facilities.

    \item We speculate as to the interior structure of super-Earths, specifically whether metal cores will, in fact, differentiate as in the case of Earth. The abundance of light species, such as hydrogen and oxygen, may affect the buoyancy and rainout of metal within the magma ocean, as also discussed in \citet{Lichtenberg2021,Schlichting2022}.
\end{itemize}

We emphasise that our calculations are meant as a first attempt to understand the chemical state of super-Earths in the context of thermal and mass loss evolution. Indeed, many approximations have been made in this work in an attempt to simplify the problem. Future work is needed on many fronts to improve our understanding of these planets. For example, coupling atmospheric evolution models with interior structure models with the necessary chemistry in a self-consistent manner will allow one to examine the sequestration process with time. Furthermore, solving for the atmospheric abundances as a function of pressure and temperature allows for more accurate predictions for atmospheric characterisation \cite[see][]{Markham2022,Misener2022,Misener2023}. Finally, more work is needed on the experimental \citep[e.g.][]{Horn2023,Piet2023} and \textit{ab initio} calculations \citep[e.g.][]{Gilmore2019} to better understand equations of state, miscibility and solubility properties of various species, such as iron, hydrogen, water, and silicates, under high pressures and temperatures.

\section*{Acknowledgements}
{\jr{We kindly thank the anonymous reviewer for comments that helped improve the paper.}} JGR is sponsored by the National Aeronautics and Space Administration (NASA) through a contract with Oak Ridge Associated Universities (ORAU). JGR was also supported by the Alfred P. Sloan Foundation under grant G202114194 as part of the AEThER collaboration. H.E.S. gratefully acknowledges NASA grant 80NSSC18K0828 for financial support during preparation and submission of the work. For the purpose of open access, the authors have applied a Creative Commons Attribution (CC-BY) licence to any Author Accepted Manuscript version arising.

\bibliography{references}{}

\begin{thebibliography}{}
\expandafter\ifx\csname natexlab\endcsname\relax\def\natexlab#1{#1}\fi
\providecommand{\url}[1]{\href{#1}{#1}}
\providecommand{\dodoi}[1]{doi:~\href{http://doi.org/#1}{\nolinkurl{#1}}}
\providecommand{\doeprint}[1]{\href{http://ascl.net/#1}{\nolinkurl{http://ascl.net/#1}}}
\providecommand{\doarXiv}[1]{\href{https://arxiv.org/abs/#1}{\nolinkurl{https://arxiv.org/abs/#1}}}

\bibitem[{{Agol} {et~al.}(2021){Agol}, {Dorn}, {Grimm}, {Turbet}, {Ducrot}, {Delrez}, {Gillon}, {Demory}, {Burdanov}, {Barkaoui}, {Benkhaldoun}, {Bolmont}, {Burgasser}, {Carey}, {de Wit}, {Fabrycky}, {Foreman-Mackey}, {Haldemann}, {Hernandez}, {Ingalls}, {Jehin}, {Langford}, {Leconte}, {Lederer}, {Luger}, {Malhotra}, {Meadows}, {Morris}, {Pozuelos}, {Queloz}, {Raymond}, {Selsis}, {Sestovic}, {Triaud}, \& {Van Grootel}}]{Agol2021}
{Agol}, E., {Dorn}, C., {Grimm}, S.~L., {et~al.} 2021, PSJ, 2, 1, \dodoi{10.3847/PSJ/abd022}

\bibitem[{{Anders} \& {Grevesse}(1989)}]{Anders1989}
{Anders}, E., \& {Grevesse}, N. 1989, \gca, 53, 197, \dodoi{10.1016/0016-7037(89)90286-X}

\bibitem[{{Anderson} {et~al.}(2001){Anderson}, {Dubrovinsky}, {Saxena}, \& {LeBihan}}]{Anderson2001}
{Anderson}, O.~L., {Dubrovinsky}, L., {Saxena}, S.~K., \& {LeBihan}, T. 2001, \grl, 28, 399, \dodoi{10.1029/2000GL008544}

\bibitem[{{Andrault} {et~al.}(2011){Andrault}, {Bolfan-Casanova}, {Nigro}, {Bouhifd}, {Garbarino}, \& {Mezouar}}]{Andrault2011}
{Andrault}, D., {Bolfan-Casanova}, N., {Nigro}, G.~L., {et~al.} 2011, Earth and Planetary Science Letters, 304, 251, \dodoi{10.1016/j.epsl.2011.02.006}

\bibitem[{{Badro} {et~al.}(2015){Badro}, {Brodholt}, {Piet}, {Siebert}, \& {Ryerson}}]{Badro2015}
{Badro}, J., {Brodholt}, J.~P., {Piet}, H., {Siebert}, J., \& {Ryerson}, F.~J. 2015, Proceedings of the National Academy of Science, 112, 12310, \dodoi{10.1073/pnas.1505672112}

\bibitem[{{Bean} {et~al.}(2021){Bean}, {Raymond}, \& {Owen}}]{Bean2021}
{Bean}, J.~L., {Raymond}, S.~N., \& {Owen}, J.~E. 2021, Journal of Geophysical Research (Planets), 126, e06639, \dodoi{10.1029/2020JE006639}

\bibitem[{{Benneke} {et~al.}(2019){Benneke}, {Knutson}, {Lothringer}, {Crossfield}, {Moses}, {Morley}, {Kreidberg}, {Fulton}, {Dragomir}, {Howard}, {Wong}, {D{\'e}sert}, {McCullough}, {Kempton}, {Fortney}, {Gilliland}, {Deming}, \& {Kammer}}]{Benneke2019}
{Benneke}, B., {Knutson}, H.~A., {Lothringer}, J., {et~al.} 2019, Nature Astronomy, 3, 813, \dodoi{10.1038/s41550-019-0800-5}

\bibitem[{{Berger} {et~al.}(2020){Berger}, {Huber}, {van Saders}, {Gaidos}, {Tayar}, \& {Kraus}}]{Berger2020b}
{Berger}, T.~A., {Huber}, D., {van Saders}, J.~L., {et~al.} 2020, \aj, 159, 280, \dodoi{10.3847/1538-3881/159/6/280}

\bibitem[{{Birch}(1964)}]{Birch1964}
{Birch}, F. 1964, \jgr, 69, 4377, \dodoi{10.1029/JZ069i020p04377}

\bibitem[{{Bower} {et~al.}(2022){Bower}, {Hakim}, {Sossi}, \& {Sanan}}]{Bower2022}
{Bower}, D.~J., {Hakim}, K., {Sossi}, P.~A., \& {Sanan}, P. 2022, \psj, 3, 93, \dodoi{10.3847/PSJ/ac5fb1}

\bibitem[{{Chachan} \& {Stevenson}(2018)}]{Chachan2018}
{Chachan}, Y., \& {Stevenson}, D.~J. 2018, \apj, 854, 21, \dodoi{10.3847/1538-4357/aaa459}

\bibitem[{{Charnoz} {et~al.}(2023){Charnoz}, {Falco}, {Tremblin}, {Sossi}, {Caracas}, \& {Lagage}}]{Charnoz2023}
{Charnoz}, S., {Falco}, A., {Tremblin}, P., {et~al.} 2023, \aap, 674, A224, \dodoi{10.1051/0004-6361/202245763}

\bibitem[{{Chassefi{\`e}re}(1996)}]{Chassefiere1996}
{Chassefi{\`e}re}, E. 1996, \jgr, 101, 26039, \dodoi{10.1029/96JE01951}

\bibitem[{{Chen} \& {Rogers}(2016)}]{ChenRogers2016}
{Chen}, H., \& {Rogers}, L.~A. 2016, \apj, 831, 180, \dodoi{10.3847/0004-637X/831/2/180}

\bibitem[{{Curry} {et~al.}(2024){Curry}, {Booth}, {Owen}, \& {Mohanty}}]{curry2024}
{Curry}, A., {Booth}, R., {Owen}, J.~E., \& {Mohanty}, S. 2024, \mnras, 528, 4314, \dodoi{10.1093/mnras/stae191}

\bibitem[{{Dorn} \& {Lichtenberg}(2021)}]{Dorn2021}
{Dorn}, C., \& {Lichtenberg}, T. 2021, \apjl, 922, L4, \dodoi{10.3847/2041-8213/ac33af}

\bibitem[{{Dos Santos}(2023)}]{DosSantos2023a}
{Dos Santos}, L.~A. 2023, IAU Symposium, 370, 56, \dodoi{10.1017/S1743921322004239}

\bibitem[{{Doyle} {et~al.}(2019){Doyle}, {Young}, {Klein}, {Zuckerman}, \& {Schlichting}}]{Doyle2019}
{Doyle}, A.~E., {Young}, E.~D., {Klein}, B., {Zuckerman}, B., \& {Schlichting}, H.~E. 2019, Science, 366, 356, \dodoi{10.1126/science.aax3901}

\bibitem[{{Foreman-Mackey} {et~al.}(2014){Foreman-Mackey}, {Hogg}, \& {Morton}}]{ForemanMackey2014}
{Foreman-Mackey}, D., {Hogg}, D.~W., \& {Morton}, T.~D. 2014, \apj, 795, 64, \dodoi{10.1088/0004-637X/795/1/64}

\bibitem[{{Fressin} {et~al.}(2013){Fressin}, {Torres}, {Charbonneau}, {Bryson}, {Christiansen}, {Dressing}, {Jenkins}, {Walkowicz}, \& {Batalha}}]{Fressin2013}
{Fressin}, F., {Torres}, G., {Charbonneau}, D., {et~al.} 2013, \apj, 766, 81, \dodoi{10.1088/0004-637X/766/2/81}

\bibitem[{{Fulton} {et~al.}(2017){Fulton}, {Petigura}, {Howard}, {Isaacson}, {Marcy}, {Cargile}, {Hebb}, {Weiss}, {Johnson}, {Morton}, {Sinukoff}, {Crossfield}, \& {Hirsch}}]{Fulton2017}
{Fulton}, B.~J., {Petigura}, E.~A., {Howard}, A.~W., {et~al.} 2017, \aj, 154, 109, \dodoi{10.3847/1538-3881/aa80eb}

\bibitem[{{Geiss}(1982)}]{Geiss1982}
{Geiss}, J. 1982, \ssr, 33, 201, \dodoi{10.1007/BF00213254}

\bibitem[{{Geiss} {et~al.}(1970){Geiss}, {Hirt}, \& {Leutwyler}}]{Geiss1970}
{Geiss}, J., {Hirt}, P., \& {Leutwyler}, H. 1970, \solphys, 12, 458, \dodoi{10.1007/BF00148028}

\bibitem[{{Gilmore} \& {Stixrude}(2019)}]{Gilmore2019}
{Gilmore}, T., \& {Stixrude}, L.~P. 2019, in AGU Fall Meeting Abstracts, Vol. 2019, MR23B--0107

\bibitem[{{Ginzburg} {et~al.}(2016){Ginzburg}, {Schlichting}, \& {Sari}}]{Ginzburg2016}
{Ginzburg}, S., {Schlichting}, H.~E., \& {Sari}, R. 2016, \apj, 825, 29, \dodoi{10.3847/0004-637X/825/1/29}

\bibitem[{{Ginzburg} {et~al.}(2018){Ginzburg}, {Schlichting}, \& {Sari}}]{Ginzburg2018}
---. 2018, \mnras, 476, 759, \dodoi{10.1093/mnras/sty290}

\bibitem[{{Goffo} {et~al.}(2023){Goffo}, {Gandolfi}, {Egger}, {Mustill}, {Albrecht}, {Hirano}, {Kochukhov}, {Astudillo-Defru}, {Barragan}, {Serrano}, {Hatzes}, {Alibert}, {Guenther}, {Dai}, {Lam}, {Csizmadia}, {Smith}, {Fossati}, {Luque}, {Rodler}, {Winther}, {R{\o}rsted}, {Alarcon}, {Bonfils}, {Cochran}, {Deeg}, {Jenkins}, {Korth}, {Livingston}, {Meech}, {Murgas}, {Orell-Miquel}, {Osborne}, {Palle}, {Persson}, {Redfield}, {Ricker}, {Seager}, {Vanderspek}, {Van Eylen}, \& {Winn}}]{Goffo2023}
{Goffo}, E., {Gandolfi}, D., {Egger}, J.~A., {et~al.} 2023, \apjl, 955, L3, \dodoi{10.3847/2041-8213/ace0c7}

\bibitem[{{Goodman} \& {Weare}(2010)}]{GoodmanWeare2010}
{Goodman}, J., \& {Weare}, J. 2010, Communications in Applied Mathematics and Computational Science, 5, 65, \dodoi{10.2140/camcos.2010.5.65}

\bibitem[{{Gupta} \& {Schlichting}(2019)}]{Gupta2019}
{Gupta}, A., \& {Schlichting}, H.~E. 2019, \mnras, 487, 24, \dodoi{10.1093/mnras/stz1230}

\bibitem[{{Gupta} \& {Schlichting}(2020)}]{Gupta2020}
---. 2020, \mnras, 493, 792, \dodoi{10.1093/mnras/staa315}

\bibitem[{{Ho} {et~al.}(2024){Ho}, {Rogers}, {Van Eylen}, {Owen}, \& {Schlichting}}]{Ho2024}
{Ho}, C. S.~K., {Rogers}, J.~G., {Van Eylen}, V., {Owen}, J.~E., \& {Schlichting}, H.~E. 2024, arXiv e-prints, arXiv:2401.12378, \dodoi{10.48550/arXiv.2401.12378}

\bibitem[{{Horn} {et~al.}(2023){Horn}, {Prakapenka}, {Chariton}, {Speziale}, \& {Shim}}]{Horn2023}
{Horn}, H.~W., {Prakapenka}, V., {Chariton}, S., {Speziale}, S., \& {Shim}, S.~H. 2023, \psj, 4, 30, \dodoi{10.3847/PSJ/acab03}

\bibitem[{{Howard} {et~al.}(2012){Howard}, {Marcy}, {Bryson}, {Jenkins}, {Rowe}, {Batalha}, {Borucki}, {Koch}, {Dunham}, {Gautier}, {Van Cleve}, {Cochran}, {Latham}, {Lissauer}, {Torres}, {Brown}, {Gilliland}, {Buchhave}, {Caldwell}, {Christensen-Dalsgaard}, {Ciardi}, {Fressin}, {Haas}, {Howell}, {Kjeldsen}, {Seager}, {Rogers}, {Sasselov}, {Steffen}, {Basri}, {Charbonneau}, {Christiansen}, {Clarke}, {Dupree}, {Fabrycky}, {Fischer}, {Ford}, {Fortney}, {Tarter}, {Girouard}, {Holman}, {Johnson}, {Klaus}, {Machalek}, {Moorhead}, {Morehead}, {Ragozzine}, {Tenenbaum}, {Twicken}, {Quinn}, {Isaacson}, {Shporer}, {Lucas}, {Walkowicz}, {Welsh}, {Boss}, {Devore}, {Gould}, {Smith}, {Morris}, {Prsa}, {Morton}, {Still}, {Thompson}, {Mullally}, {Endl}, \& {MacQueen}}]{Howard2012}
{Howard}, A.~W., {Marcy}, G.~W., {Bryson}, S.~T., {et~al.} 2012, \apjs, 201, 15, \dodoi{10.1088/0067-0049/201/2/15}

\bibitem[{{Hunten} {et~al.}(1987){Hunten}, {Pepin}, \& {Walker}}]{Hunten1987}
{Hunten}, D.~M., {Pepin}, R.~O., \& {Walker}, J.~C.~G. 1987, \icarus, 69, 532, \dodoi{10.1016/0019-1035(87)90022-4}

\bibitem[{{Jontof-Hutter} {et~al.}(2014){Jontof-Hutter}, {Lissauer}, {Rowe}, \& {Fabrycky}}]{JontofHutter2014}
{Jontof-Hutter}, D., {Lissauer}, J.~J., {Rowe}, J.~F., \& {Fabrycky}, D.~C. 2014, \apj, 785, 15, \dodoi{10.1088/0004-637X/785/1/15}

\bibitem[{{Joselyn} \& {Holzer}(1978)}]{Joselyn1978}
{Joselyn}, J., \& {Holzer}, T.~E. 1978, \jgr, 83, 1019, \dodoi{10.1029/JA083iA03p01019}

\bibitem[{{Karki} {et~al.}(2000){Karki}, {Wentzcovitch}, {de Gironcoli}, \& {Baroni}}]{Karki2000}
{Karki}, B.~B., {Wentzcovitch}, R.~M., {de Gironcoli}, S., \& {Baroni}, S. 2000, \prb, 62, 14750, \dodoi{10.1103/PhysRevB.62.14750}

\bibitem[{{Kite} {et~al.}(2019){Kite}, {Fegley}, {Schaefer}, \& {Ford}}]{Kite2019}
{Kite}, E.~S., {Fegley}, Bruce, J., {Schaefer}, L., \& {Ford}, E.~B. 2019, \apjl, 887, L33, \dodoi{10.3847/2041-8213/ab59d9}

\bibitem[{{Kite} {et~al.}(2020){Kite}, {Fegley}, {Schaefer}, \& {Ford}}]{Kite2020}
---. 2020, \apj, 891, 111, \dodoi{10.3847/1538-4357/ab6ffb}

\bibitem[{{Kite} {et~al.}(2016){Kite}, {Fegley}, {Schaefer}, \& {Gaidos}}]{Kite2016}
{Kite}, E.~S., {Fegley}, Bruce, J., {Schaefer}, L., \& {Gaidos}, E. 2016, \apj, 828, 80, \dodoi{10.3847/0004-637X/828/2/80}

\bibitem[{{Kite} \& {Schaefer}(2021)}]{Kite2021}
{Kite}, E.~S., \& {Schaefer}, L. 2021, \apjl, 909, L22, \dodoi{10.3847/2041-8213/abe7dc}

\bibitem[{{Kubyshkina} \& {Fossati}(2022)}]{Kubyshkina2022}
{Kubyshkina}, D., \& {Fossati}, L. 2022, arXiv e-prints, arXiv:2211.10166.
\newblock \doarXiv{2211.10166}

\bibitem[{{Lam} {et~al.}(2021){Lam}, {Csizmadia}, {Astudillo-Defru}, {Bonfils}, {Gandolfi}, {Padovan}, {Esposito}, {Hellier}, {Hirano}, {Livingston}, {Murgas}, {Smith}, {Collins}, {Mathur}, {Garcia}, {Howell}, {Santos}, {Dai}, {Ricker}, {Vanderspek}, {Latham}, {Seager}, {Winn}, {Jenkins}, {Albrecht}, {Almenara}, {Artigau}, {Barrag{\'a}n}, {Bouchy}, {Cabrera}, {Charbonneau}, {Chaturvedi}, {Chaushev}, {Christiansen}, {Cochran}, {De Meideiros}, {Delfosse}, {D{\'\i}az}, {Doyon}, {Eigm{\"u}ller}, {Figueira}, {Forveille}, {Fridlund}, {Gaisn{\'e}}, {Goffo}, {Georgieva}, {Grziwa}, {Guenther}, {Hatzes}, {Johnson}, {Kab{\'a}th}, {Knudstrup}, {Korth}, {Lewin}, {Lissauer}, {Lovis}, {Luque}, {Melo}, {Morgan}, {Morris}, {Mayor}, {Narita}, {Osborne}, {Palle}, {Pepe}, {Persson}, {Quinn}, {Rauer}, {Redfield}, {Schlieder}, {S{\'e}gransan}, {Serrano}, {Smith}, {{\v{S}}ubjak}, {Twicken}, {Udry}, {Van Eylen}, \& {Vezie}}]{Lam2021}
{Lam}, K. W.~F., {Csizmadia}, S., {Astudillo-Defru}, N., {et~al.} 2021, Science, 374, 1271, \dodoi{10.1126/science.aay3253}

\bibitem[{{Lee} {et~al.}(2014){Lee}, {Chiang}, \& {Ormel}}]{Lee2014}
{Lee}, E.~J., {Chiang}, E., \& {Ormel}, C.~W. 2014, \apj, 797, 95, \dodoi{10.1088/0004-637X/797/2/95}

\bibitem[{{Lichtenberg}(2021)}]{Lichtenberg2021}
{Lichtenberg}, T. 2021, \apjl, 914, L4, \dodoi{10.3847/2041-8213/ac0146}

\bibitem[{{Lopez} \& {Fortney}(2013)}]{LopezFortney2013}
{Lopez}, E.~D., \& {Fortney}, J.~J. 2013, \apj, 776, 2, \dodoi{10.1088/0004-637X/776/1/2}

\bibitem[{{Lopez} \& {Fortney}(2014)}]{Lopez2014}
---. 2014, \apj, 792, 1, \dodoi{10.1088/0004-637X/792/1/1}

\bibitem[{{Luger} \& {Barnes}(2015)}]{Luger2015}
{Luger}, R., \& {Barnes}, R. 2015, Astrobiology, 15, 119, \dodoi{10.1089/ast.2014.1231}

\bibitem[{{Madhusudhan} {et~al.}(2023){Madhusudhan}, {Sarkar}, {Constantinou}, {Holmberg}, {Piette}, \& {Moses}}]{Madhusudhan2023}
{Madhusudhan}, N., {Sarkar}, S., {Constantinou}, S., {et~al.} 2023, \apjl, 956, L13, \dodoi{10.3847/2041-8213/acf577}

\bibitem[{{Markham} {et~al.}(2022){Markham}, {Guillot}, \& {Stevenson}}]{Markham2022}
{Markham}, S., {Guillot}, T., \& {Stevenson}, D. 2022, \aap, 665, A12, \dodoi{10.1051/0004-6361/202243359}

\bibitem[{{Misener} \& {Schlichting}(2021)}]{Misener2021}
{Misener}, W., \& {Schlichting}, H.~E. 2021, \mnras, 503, 5658, \dodoi{10.1093/mnras/stab895}

\bibitem[{{Misener} \& {Schlichting}(2022)}]{Misener2022}
---. 2022, \mnras, 514, 6025, \dodoi{10.1093/mnras/stac1732}

\bibitem[{{Misener} {et~al.}(2023){Misener}, {Schlichting}, \& {Young}}]{Misener2023}
{Misener}, W., {Schlichting}, H.~E., \& {Young}, E.~D. 2023, \mnras, 524, 981, \dodoi{10.1093/mnras/stad1910}

\bibitem[{{Moore} {et~al.}(1998){Moore}, {Vennemann}, \& {Carmichael}}]{Moore1998}
{Moore}, G., {Vennemann}, T., \& {Carmichael}, I.~S.~E. 1998, American Mineralogist, 83, 36, \dodoi{10.2138/am-1998-1-203}

\bibitem[{{Mulders} {et~al.}(2018){Mulders}, {Pascucci}, {Apai}, \& {Ciesla}}]{Mulders2018}
{Mulders}, G.~D., {Pascucci}, I., {Apai}, D., \& {Ciesla}, F.~J. 2018, \aj, 156, 24, \dodoi{10.3847/1538-3881/aac5ea}

\bibitem[{{Murray-Clay} {et~al.}(2009){Murray-Clay}, {Chiang}, \& {Murray}}]{MurrayClay2009}
{Murray-Clay}, R.~A., {Chiang}, E.~I., \& {Murray}, N. 2009, \apj, 693, 23, \dodoi{10.1088/0004-637X/693/1/23}

\bibitem[{{Olson} \& {Sharp}(2018)}]{Olson2018}
{Olson}, P., \& {Sharp}, Z.~D. 2018, Earth and Planetary Science Letters, 498, 418, \dodoi{10.1016/j.epsl.2018.07.006}

\bibitem[{{Otegi} {et~al.}(2020){Otegi}, {Bouchy}, \& {Helled}}]{Otegi2020}
{Otegi}, J.~F., {Bouchy}, F., \& {Helled}, R. 2020, \aap, 634, A43, \dodoi{10.1051/0004-6361/201936482}

\bibitem[{{Owen} \& {Campos Estrada}(2020)}]{OwenCamposEstrada2020}
{Owen}, J.~E., \& {Campos Estrada}, B. 2020, \mnras, 491, 5287, \dodoi{10.1093/mnras/stz3435}

\bibitem[{{Owen} \& {Jackson}(2012)}]{OwenJackson2012}
{Owen}, J.~E., \& {Jackson}, A.~P. 2012, \mnras, 425, 2931, \dodoi{10.1111/j.1365-2966.2012.21481.x}

\bibitem[{{Owen} \& {Lai}(2018)}]{OwenLai2018}
{Owen}, J.~E., \& {Lai}, D. 2018, \mnras, 479, 5012, \dodoi{10.1093/mnras/sty1760}

\bibitem[{{Owen} \& {Schlichting}(2023)}]{OwenSchlichting2023}
{Owen}, J.~E., \& {Schlichting}, H.~E. 2023, \mnras, \dodoi{10.1093/mnras/stad3972}

\bibitem[{{Owen} \& {Wu}(2013)}]{Owen2013}
{Owen}, J.~E., \& {Wu}, Y. 2013, \apj, 775, 105, \dodoi{10.1088/0004-637X/775/2/105}

\bibitem[{{Owen} \& {Wu}(2016)}]{Owen2016}
---. 2016, \apj, 817, 107, \dodoi{10.3847/0004-637X/817/2/107}

\bibitem[{{Owen} \& {Wu}(2017)}]{Owen2017}
---. 2017, \apj, 847, 29, \dodoi{10.3847/1538-4357/aa890a}

\bibitem[{{Petigura} {et~al.}(2022){Petigura}, {Rogers}, {Isaacson}, {Owen}, {Kraus}, {Winn}, {MacDougall}, {Howard}, {Fulton}, {Kosiarek}, {Weiss}, {Behmard}, \& {Blunt}}]{Petigura2022}
{Petigura}, E.~A., {Rogers}, J.~G., {Isaacson}, H., {et~al.} 2022, \aj, 163, 179, \dodoi{10.3847/1538-3881/ac51e3}

\bibitem[{{Piet} {et~al.}(2023){Piet}, {Chizmeshya}, {Chen}, {Chariton}, {Greenberg}, {Prakapenka}, {Buseck}, \& {Shim}}]{Piet2023}
{Piet}, H., {Chizmeshya}, A., {Chen}, B., {et~al.} 2023, \grl, 50, e2022GL101155, \dodoi{10.1029/2022GL101155}

\bibitem[{{Rafikov}(2006)}]{Rafikov2006}
{Rafikov}, R.~R. 2006, \apj, 648, 666, \dodoi{10.1086/505695}

\bibitem[{{Rogers} {et~al.}(2021){Rogers}, {Gupta}, {Owen}, \& {Schlichting}}]{Rogers2021b}
{Rogers}, J.~G., {Gupta}, A., {Owen}, J.~E., \& {Schlichting}, H.~E. 2021, \mnras, 508, 5886, \dodoi{10.1093/mnras/stab2897}

\bibitem[{{Rogers} {et~al.}(2023{\natexlab{a}}){Rogers}, {Jan{\'o} Mu{\~n}oz}, {Owen}, \& {Makinen}}]{Rogers2023}
{Rogers}, J.~G., {Jan{\'o} Mu{\~n}oz}, C., {Owen}, J.~E., \& {Makinen}, T.~L. 2023{\natexlab{a}}, \mnras, 519, 6028, \dodoi{10.1093/mnras/stad089}

\bibitem[{{Rogers} \& {Owen}(2021)}]{Rogers2021}
{Rogers}, J.~G., \& {Owen}, J.~E. 2021, \mnras, 503, 1526, \dodoi{10.1093/mnras/stab529}

\bibitem[{{Rogers} {et~al.}(2024){Rogers}, {Owen}, \& {Schlichting}}]{Rogers2024}
{Rogers}, J.~G., {Owen}, J.~E., \& {Schlichting}, H.~E. 2024, \mnras, 529, 2716, \dodoi{10.1093/mnras/stae563}

\bibitem[{{Rogers} {et~al.}(2023{\natexlab{b}}){Rogers}, {Schlichting}, \& {Owen}}]{Rogers2023b}
{Rogers}, J.~G., {Schlichting}, H.~E., \& {Owen}, J.~E. 2023{\natexlab{b}}, \apjl, 947, L19, \dodoi{10.3847/2041-8213/acc86f}

\bibitem[{{Rogers}(2015)}]{Rogers2015}
{Rogers}, L.~A. 2015, \apj, 801, 41, \dodoi{10.1088/0004-637X/801/1/41}

\bibitem[{{Rogers} \& {Seager}(2010)}]{Rogers2010}
{Rogers}, L.~A., \& {Seager}, S. 2010, \apj, 712, 974, \dodoi{10.1088/0004-637X/712/2/974}

\bibitem[{{Salvador} \& {Samuel}(2023)}]{Salvador2023}
{Salvador}, A., \& {Samuel}, H. 2023, \icarus, 390, 115265, \dodoi{10.1016/j.icarus.2022.115265}

\bibitem[{{Schlichting} \& {Young}(2022)}]{Schlichting2022}
{Schlichting}, H.~E., \& {Young}, E.~D. 2022, PSJ, 3, 127, \dodoi{10.3847/PSJ/ac68e6}

\bibitem[{{Schulik} \& {Booth}(2022)}]{Schulik2022}
{Schulik}, M., \& {Booth}, R. 2022, arXiv e-prints, arXiv:2207.07144.
\newblock \doarXiv{2207.07144}

\bibitem[{{Seager} {et~al.}(2007){Seager}, {Kuchner}, {Hier-Majumder}, \& {Militzer}}]{Seager2007}
{Seager}, S., {Kuchner}, M., {Hier-Majumder}, C.~A., \& {Militzer}, B. 2007, \apj, 669, 1279, \dodoi{10.1086/521346}

\bibitem[{{Silburt} {et~al.}(2015){Silburt}, {Gaidos}, \& {Wu}}]{Silburt2015}
{Silburt}, A., {Gaidos}, E., \& {Wu}, Y. 2015, \apj, 799, 180, \dodoi{10.1088/0004-637X/799/2/180}

\bibitem[{{Stixrude}(2014)}]{Stixrude2014}
{Stixrude}, L. 2014, Philosophical Transactions of the Royal Society of London Series A, 372, 20130076, \dodoi{10.1098/rsta.2013.0076}

\bibitem[{{Suer} {et~al.}(2023){Suer}, {Jackson}, {Grewal}, {Dalou}, \& {Lichtenberg}}]{Suer2023}
{Suer}, T.-A., {Jackson}, C., {Grewal}, D.~S., {Dalou}, C., \& {Lichtenberg}, T. 2023, Frontiers in Earth Science, 11, 1159412, \dodoi{10.3389/feart.2023.1159412}

\bibitem[{{Trierweiler} {et~al.}(2023){Trierweiler}, {Doyle}, \& {Young}}]{Trierweiler2023}
{Trierweiler}, I.~L., {Doyle}, A.~E., \& {Young}, E.~D. 2023, \psj, 4, 136, \dodoi{10.3847/PSJ/acdef3}

\bibitem[{{Van Eylen} {et~al.}(2018){Van Eylen}, {Agentoft}, {Lundkvist}, {Kjeldsen}, {Owen}, {Fulton}, {Petigura}, \& {Snellen}}]{VanEylen2018}
{Van Eylen}, V., {Agentoft}, C., {Lundkvist}, M.~S., {et~al.} 2018, \mnras, 479, 4786, \dodoi{10.1093/mnras/sty1783}

\bibitem[{Virtanen {et~al.}(2020)Virtanen, Gommers, Oliphant, Haberland, Reddy, Cournapeau, Burovski, Peterson, Weckesser, Bright, {van der Walt}, Brett, Wilson, Millman, Mayorov, Nelson, Jones, Kern, Larson, Carey, Polat, Feng, Moore, {VanderPlas}, Laxalde, Perktold, Cimrman, Henriksen, Quintero, Harris, Archibald, Ribeiro, Pedregosa, {van Mulbregt}, \& {SciPy 1.0 Contributors}}]{SciPy2020}
Virtanen, P., Gommers, R., Oliphant, T.~E., {et~al.} 2020, Nature Methods, 17, 261, \dodoi{10.1038/s41592-019-0686-2}

\bibitem[{{Weiss} \& {Marcy}(2014)}]{Weiss2014}
{Weiss}, L.~M., \& {Marcy}, G.~W. 2014, \apj, 783, L6, \dodoi{10.1088/2041-8205/783/1/L6}

\bibitem[{{Wogan} {et~al.}(2024){Wogan}, {Batalha}, {Zahnle}, {Krissansen-Totton}, {Tsai}, \& {Hu}}]{Wogan2024}
{Wogan}, N.~F., {Batalha}, N.~E., {Zahnle}, K., {et~al.} 2024, arXiv e-prints, arXiv:2401.11082, \dodoi{10.48550/arXiv.2401.11082}

\bibitem[{{Wood}(2008)}]{Wood2008}
{Wood}, B.~J. 2008, Philosophical Transactions of the Royal Society of London Series A, 366, 4339, \dodoi{10.1098/rsta.2008.0115}

\bibitem[{{Wu}(2019)}]{Wu2019}
{Wu}, Y. 2019, \apj, 874, 91, \dodoi{10.3847/1538-4357/ab06f8}

\bibitem[{{Xiang} {et~al.}(1997){Xiang}, {Sun}, {Fan}, \& {Gong}}]{Xiang1997}
{Xiang}, Y., {Sun}, D.~Y., {Fan}, W., \& {Gong}, X.~G. 1997, Physics Letters A, 233, 216, \dodoi{10.1016/S0375-9601(97)00474-X}

\bibitem[{{Young} {et~al.}(2023){Young}, {Shahar}, \& {Schlichting}}]{Young2023}
{Young}, E.~D., {Shahar}, A., \& {Schlichting}, H.~E. 2023, \nat, 616, 306, \dodoi{10.1038/s41586-023-05823-0}

\bibitem[{{Zahnle} \& {Kasting}(1986)}]{Zahnle1986}
{Zahnle}, K.~J., \& {Kasting}, J.~F. 1986, \icarus, 68, 462, \dodoi{10.1016/0019-1035(86)90051-5}

\bibitem[{{Zhang} \& {Rogers}(2022)}]{J_Zhang2022}
{Zhang}, J., \& {Rogers}, L.~A. 2022, \apj, 938, 131, \dodoi{10.3847/1538-4357/ac8e65}

\bibitem[{{Zink} {et~al.}(2019){Zink}, {Christiansen}, \& {Hansen}}]{Zink2019}
{Zink}, J.~K., {Christiansen}, J.~L., \& {Hansen}, B. M.~S. 2019, \mnras, 483, 4479, \dodoi{10.1093/mnras/sty3463}

\end{thebibliography}
\bibliographystyle{aasjournal}

\appendix

\section{Chemical Reactions} \label{app:reactions}
Our chemical equilibrium model relies on a set of linearly independent reactions, which are discussed in detail in \citet{Schlichting2022}. Here, we list the reaction set, which spans our reaction space, allowing for all available pathways for our chosen species. Firstly, speciation within the magma ocean:

\begin{align*} 
    \text{Na}_2 \text{Si} \text{O}_3 \;\; & \rightleftharpoons \;\; \text{Na}_2  \text{O} + \text{SiO}_2, \tag{A1} \\[0.3cm]
    \text{MgSiO}_3  \;\; & \rightleftharpoons \;\; \text{MgO} + \text{SiO}_2, \tag{A2} \\[0.3cm]
    \text{FeSiO}_3  \;\; & \rightleftharpoons \;\; \text{FeO} + \text{SiO}_2, \tag{A3} \\[0.3cm]
    2\text{H}_\text{metal}  \;\; & \rightleftharpoons \;\; \text{H}_\text{2,silicate}, \tag{A4} \\[0.3cm]
    2\text{H}_2 \text{O}_\text{silicate} +\text{Si}_\text{metal} \;\; & \rightleftharpoons \;\; \text{SiO}_\text{2} + 2\text{H}_\text{2,silicate}, \tag{A5} \\[0.1cm]
    \text{O}_\text{metal}  + \frac{1}{2} \text{Si}_\text{metal} \;\;
    & \rightleftharpoons \;\; \frac{1}{2} \text{SiO}_2, \tag{A6} \\[0.1cm]
    \frac{1}{2} \text{Si} \text{O}_2  + \text{Fe}_\text{metal} \;\;
    & \rightleftharpoons \;\; \text{FeO} + \frac{1}{2} \text{Si}_\text{metal} \tag{72}. \\
\end{align*}

Then, reactions within the atmosphere:

\begin{align*} 
    \text{CO}_\text{gas} + \frac{1}{2} \text{O}_\text{2,gas} \;\; & \rightleftharpoons \;\; \text{CO}_\text{2,gas}, \tag{A8} \\[0.1cm]
    \text{CH}_\text{4,gas} + \frac{1}{2} \text{O}_\text{2,gas}, \;\; & \rightleftharpoons \;\; 2\text{H}_\text{2,gas} + \text{CO}_\text{gas}, \tag{A9} \\[0.1cm]
    \text{H}_\text{2,gas} + \frac{1}{2} \text{O}_\text{2,gas} \;\; & \rightleftharpoons \;\; \text{H}_\text{2}\text{O}_\text{gas} \tag{A10}. \\
\end{align*}

Finally, reactions between the atmosphere and magma ocean:

\begin{align*} 
    \text{FeO} \;\; & \rightleftharpoons \;\; \text{Fe}_\text{gas} + \frac{1}{2} \text{O}_\text{2,gas} \tag{A11}, \\[0.1cm]
    \text{MgO} \;\; & \rightleftharpoons \;\; \text{Mg}_\text{gas} + \frac{1}{2} \text{O}_\text{2,gas} \tag{A12}, \\[0.1cm]
    \text{SiO}_2 \;\; & \rightleftharpoons \;\; \text{SiO}_\text{gas} + \frac{1}{2} \text{O}_\text{2,gas} \tag{A13}, \\[0.1cm]
    \text{Na}_2 \text{O} \;\; & \rightleftharpoons \;\; 2\text{Na}_\text{gas} + \frac{1}{2} \text{O}_\text{2,gas} \tag{A14}, \\[0.2cm]
    \text{H}_\text{2,gas} \;\; & \rightleftharpoons \;\; \text{H}_\text{2,silicate} \tag{A15}, \\[0.3cm]
    \text{H}_2 \text{O}_\text{gas} \;\; & \rightleftharpoons \;\; \text{H}_\text{2} \text{O}_\text{silicate} \tag{A16}, \\[0.3cm]
    \text{CO}_\text{gas} \;\; & \rightleftharpoons \;\; \text{CO}_\text{silicate} \tag{A17}, \\[0.3cm]
    \text{CO}_\text{2,gas} \;\; & \rightleftharpoons \;\; \text{CO}_\text{2,silicate} \tag{A18}. \\[0.3cm]
\end{align*}




\end{document}